\documentclass[conference]{IEEEtran}

\usepackage{cite}
\usepackage{amsmath,amssymb,amsfonts}
\usepackage{algorithmic}
\usepackage{graphicx}
\usepackage{textcomp}
\usepackage{xcolor}
\usepackage{array,multirow}
\usepackage{tablefootnote}
\usepackage{threeparttable}
\usepackage{relsize}
\usepackage{enumitem}
\usepackage{amsmath}
\usepackage{upgreek}
\usepackage{url}
\usepackage[caption=false, font=footnotesize]{subfig}

\def\BibTeX{{\rm B\kern-.05em{\sc i\kern-.025em b}\kern-.08em
    T\kern-.1667em\lower.7ex\hbox{E}\kern-.125emX}}

\makeatletter
\IEEEtriggercmd{\reset@font\normalfont\fontsize{7.9pt}{8.60pt}\selectfont}
\newcommand*\titleheader[1]{\gdef\@titleheader{#1}}
\AtBeginDocument{%
  \let\st@red@title\@title
  \def\@title{%
    \bgroup\normalfont\normalsize\centering\@titleheader\par\egroup
    \vskip1ex\st@red@title}
}
\makeatother
\IEEEtriggeratref{1}

\titleheader{\color{gray}\vspace{-30pt}\hspace{-3pt}Presented at the 30th IFIP/IEEE Int'l. Conference on Very Large Scale Integration (VLSI-SoC), 03/10--05/10 2022, Patras, GR.\\This is the authors’ version of the paper. The final published version is posted on IEEE Xplore.\\ DOI: \url{https://doi.org/10.1109/VLSI-SoC54400.2022.9939621}}
  
\title{Combining Fault Tolerance Techniques and COTS SoC Accelerators for Payload Processing in Space}

\makeatletter
\def\ps@IEEEtitlepagestyle{
  \def\@oddfoot{\mycopyrightnotice}
  \def\@evenfoot{}
}
\def\mycopyrightnotice{
  {\footnotesize
  \begin{minipage}{\textwidth}
  \centering\color{gray}%
  ~\copyright~2022 IEEE.  Personal use of this material is permitted.  Permission from IEEE must be obtained for all other uses, in any current or future media, including reprinting/republishing this material for advertising or promotional purposes, creating new collective works, for resale or redistribution\\to servers or lists, or reuse of any copyrighted component of this work in other works.
  \end{minipage}
  }
}
  
\begin{document}

\author{\IEEEauthorblockN{Vasileios Leon\IEEEauthorrefmark{1}, 
    Elissaios Alexios Papatheofanous\IEEEauthorrefmark{2}, 
    George Lentaris\IEEEauthorrefmark{1},\\
    Charalampos Bezaitis\IEEEauthorrefmark{2},
    Nikolaos Mastorakis\IEEEauthorrefmark{1},
    Georgios Bampilis\IEEEauthorrefmark{1},
    Dionysios Reisis\IEEEauthorrefmark{2},
    Dimitrios Soudris\IEEEauthorrefmark{1}}\\[-11pt]
    \IEEEauthorblockA{\IEEEauthorrefmark{1}\emph{School of Electrical and Computer Engineering, National Technical University of Athens, 15780 Athens, Greece}}
    \IEEEauthorblockA{\IEEEauthorrefmark{2}\emph{Department of Physics, National and Kapodistrian University of Athens, 15772 Athens, Greece}\\}
    }
    
\maketitle

\begin{abstract}
The ever-increasing demand for computational power and I/O throughput in space applications is transforming the landscape of on-board computing. A variety of Commercial-Off-The-Shelf (COTS) accelerators emerges as an attractive solution for payload processing to outperform the traditional radiation-hardened devices. Towards increasing the reliability of such COTS accelerators, the current paper explores and evaluates fault-tolerance techniques for the Zynq FPGA and the Myriad VPU, which are two device families being integrated in industrial space avionics architectures/boards, such as Ubotica's CogniSat, Xiphos' Q7S, and Cobham Gaisler's GR-VPX-XCKU060. On the FPGA side, we combine techniques such as memory scrubbing, partial reconfiguration, triple modular redundancy, and watchdogs. On the VPU side, we detect and correct errors in the instruction and data memories, as well as we apply redundancy at processor level (SHAVE cores). When considering FPGA with VPU co-processing, we also develop a fault-tolerant interface between the two devices based on the CIF/LCD protocols and our custom CRC error-detecting code.
\end{abstract}


\section{Introduction}

The ``NewSpace'' era relies on novel technological 
approaches in space applications.
The advances in small form factor satellites, 
such as SmallSats and CubeSats, 
broaden the scope of the Earth Observation (EO) missions 
and attract new areas of research for payload processing in space.
In particular, challenges such as the increased  
data generated by
satellite instruments  
and computational demands 
of modern Digital Signal Processing (DSP) 
and Artificial Intelligence (AI) algorithms,
constrained by the power budget and  
dependability requirements of each space mission,
drives the industry to revisit the computing architectures for space avionics. 

Classical space-qualified general-purpose processors, 
such as the radiation-hardened 
Power-PC and LEON CPUs \cite{access},
fail to follow the aforementioned trend.
As a result,
the industry turns to mixed-criticality architectures. 
These heterogeneous architectures combine rad-hard components and Commercial-Off-The-Shelf (COTS) accelerators for payload processing and especially image processing applications that are not mission-critical \cite{Furano2020}.  
Specialized COTS System-on-Chip (SoC), such as Vision Processing Units (VPUs) and SoC Field-Programmable Gate Arrays (FPGAs), offer attractive trade-offs between Size, Weight, Power and Cost (SWaP-C), processing performance and development flexibility \cite{lentarisTVID, leotome}.
However, 
despite their benefits,
these COTS devices are not radiation hardened by design, 
and thus, 
they are susceptible to ionizing radiation 
increasing the risk of failures. 
To tackle this risk, 
Fault-Tolerance (FT) architectures and fault-mitigation techniques 
need to be designed. 
Commonly used techniques include rad-hard components supervising the COTS ones, Error Correction Coding (ECC) and Error Detection and Correction (EDAC) in memories, 
and Triple Modular Redundant (TMR) processing. 

Recent works in the literature showcase combinations of fault-mitigation techniques at different levels of computing architectures for space applications \cite{OBDP_Navarro, OBDP_Amorim, OBDP_Kuligowski}. 
In \cite{OBDP_Navarro}, 
platform-level FT is developed, 
where 3 distinct Myriad2 VPUs operate in TMR mode.
In this architecture, 
the Cobham Gaisler's GR-VPX-XCKU060 board \cite{vpx}
integrates
Xilinx's Kintex UltraScale XCKU060 FPGA for voting 
and the radiation-tolerant GR716 microcontroller acting as supervisor. 
In \cite{OBDP_Amorim}, 
the authors utilize a COTS Xilinx Zynq MPSoC for payload processing,
which is partitioned between isolated secure/non-secure areas based on the criticality of the application executed on them, while a rad-hard PolarFire FPGA supervises.
The work in \cite{OBDP_Kuligowski} introduces hardening techniques in a system with two redundant Zynq UltraScale+ MPSoCs for accelerating neural-network-based hyperspectral image
segmentation. 
The supervisor is a Vorago Cortex-M0 microcontroller,
which also implements EDAC/ECC in all internal memories.

The current work explores, combines, and evaluates fault-mitigation techniques for a COTS-based mixed-criticality architecture
targeting the acceleration of payload functions.
Our HW architecture consists of a SoC FPGA and Myriad2 VPU pair
with diverse implementations,
which vary from interconnecting low-end readily-available COTS boards
to developing high-end custom-made PCBs 
(depending on mission requirements and budget availability).
Our SW layer, which is the focus of this paper, 
consists of a combination of techniques
trading off reliability and performance overheads
(also depending on mission specifications).
On the FPGA side,
we implement mitigation techniques for the FPGA configuration memory and process, 
as well as modular redundancy for the payload processing acceleration. 
On the VPU side,
we apply error correction to both instruction and working memory 
as well as modular redundancy of processing cores. 
Furthermore, we develop FT communication for the payload data transfers between FPGA and VPU via 
the CIF and LCD interfaces. 
Finally, we design a supervising scheme, 
where each one device acts as an individual watchdog of the second device, 
while monitoring its status over UART. 
We test an engineering model of the proposed system 
by utilizing simple COTS boards, i.e.,
a Xilinx Zynq-7000 SoC FPGA and the Intel Myriad2 VPU,
where all of our mitigation techniques are evaluated. 
For the evaluation of our techniques, 
we adopt 
a fault emulation/injection setup.
The results show that
our robust FT architecture on Zynq
maintains correct functionality
for the 72\% of the time even during
an extreme test involving 10K injections per 40s.
On the Myriad2 side,
we ensure error-free application results
with an overhead of $\sim$50ms.

\section{COTS-based FPGA \& VPU Architecture}

\begin{figure}[!t]
    \centering
    \includegraphics[width=0.8\columnwidth]{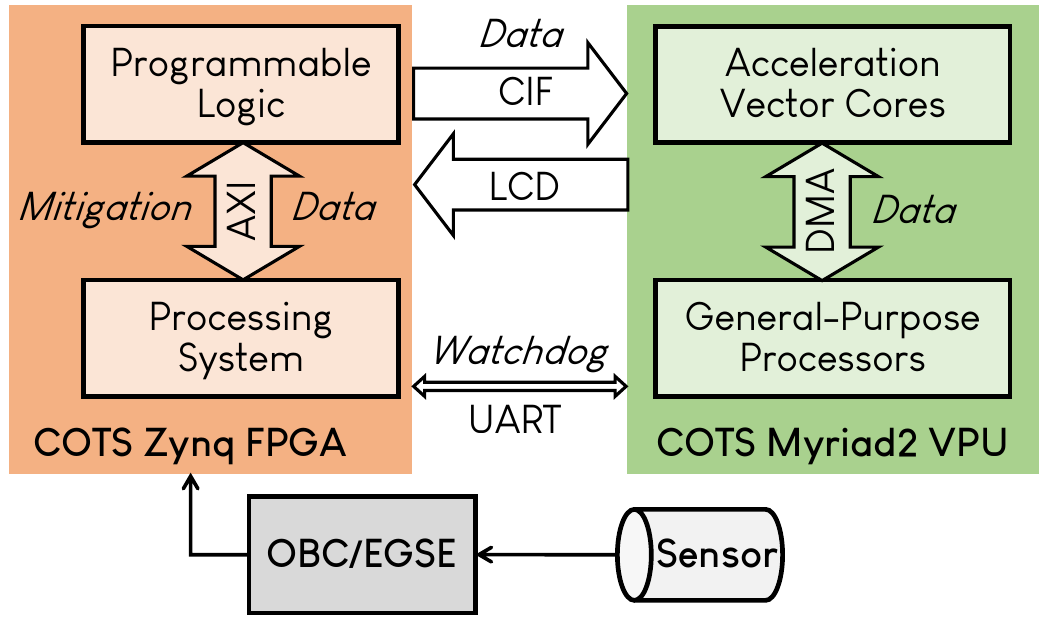}
    \vspace*{-5pt}
    \caption{High-level COTS co-processing architecture with Zynq and Myriad2.}
    \label{archi1}
    \vspace{-10pt}
\end{figure}

As a proof-of-concept, 
we assume here the instantiation of the FPGA \hspace*{-1pt}\&\hspace*{-1pt} VPU architecture
depicted in Fig. \ref{archi1}, 
which enables us to study  
our SW layer of mitigation techniques.
Flight models can rely on more sophisticated HW configurations
and PCBs.
More precisely,
depending on mission specifics,
we identify three potential HW implementation levels 
in an ascending order of development cost and tolerance: 
\begin{itemize}
    \item Xilinx Zybo FPGA + Ubotica CogniSat or EoT VPU
    \item Xiphos Q7S (with daughter card) + GR-HPCB-FMC-M2
    \item custom PCB with redundant FPGA and/or VPU units 
\end{itemize}
The first level involves relatively ubiquitous boards 
and targets low-cost CubeSats.
For more advanced CubeSats,
the second level involves commercial 
space-oriented boards,
e.g., Xiphos Q7S \cite{q7} with extra FMC adaptor card
enabling the 
Cobham Gaisler GR-HPCB-FMC-M2 daughterboard \cite{OBDP_Navarro} to 
connect and maximize the FPGA-VPU bandwidth.
The third level involves customized PCB designs   
targeting even microsatellites, 
e.g., similarly to our previous work \cite{ICECS_HPCB} 
(but with Zynq). 

The current paper 
assumes the first implementation level (Zybo plus 
CogniSat \cite{cognisat} or 
EoT \cite{eott})
with the FPGA handling the payload sensors/instruments  
(Fig. \ref{archi1}),
e.g., via SpaceWire as in our previous work \cite{leotome}.
The FPGA performs mostly data transcoding and DSP acceleration,
while it also encodes and forwards instrument data to the VPU via CIF.
The VPU mostly 
accelerates DSP/AI workloads.
The VPU results are received by the FPGA via UART. 
The boards' interconnection
relies on available pins, e.g., on CogniSat's hardwired CIF pins and Zybo's GPIOs for CIF. 
The UART interface  
can also facilitate our watchdog policies.
Boards can be externally reset via dedicated pins.
We note that CIF/LCD FPGA--VPU interconnections were also developed 
in our previous work \cite{ICECS_HPCB}.

\begin{figure*}[!t]
    \centering
    \includegraphics[width=\textwidth]{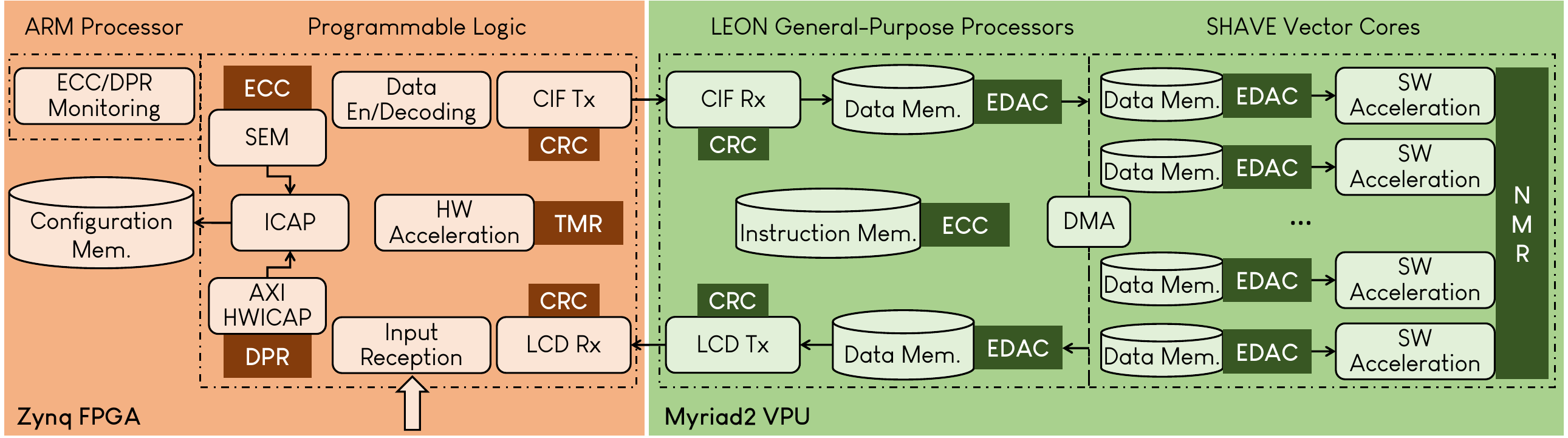}
     \vspace*{-15pt}
    \caption{Main components and dataflow of FT co-processing architecture based on Zynq and Myriad2.}
    \label{archi2}
    \vspace{-14pt}
\end{figure*}

\section{Mitigation Techniques on Zynq FPGA}

In the Zynq FPGA (left side of Fig. \ref{archi2}),
the ARM-based Processing System (PS)
is used for 
monitoring the FT techniques,
while the Programmable Logic (PL)
is used for implementing the accelerators
and all the required FT components.

\begin{figure*}[!t]
\centering
\subfloat[CMS\label{z1}]{\includegraphics[width=0.33\textwidth]{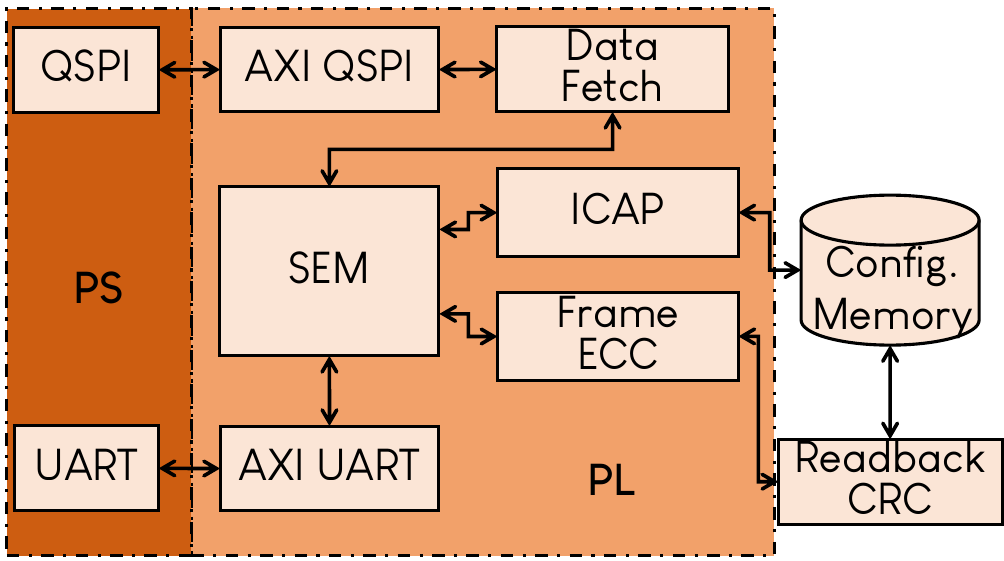}}\hspace{0.5pt}
\subfloat[DPR\label{z2}]{\includegraphics[width=0.33\textwidth]{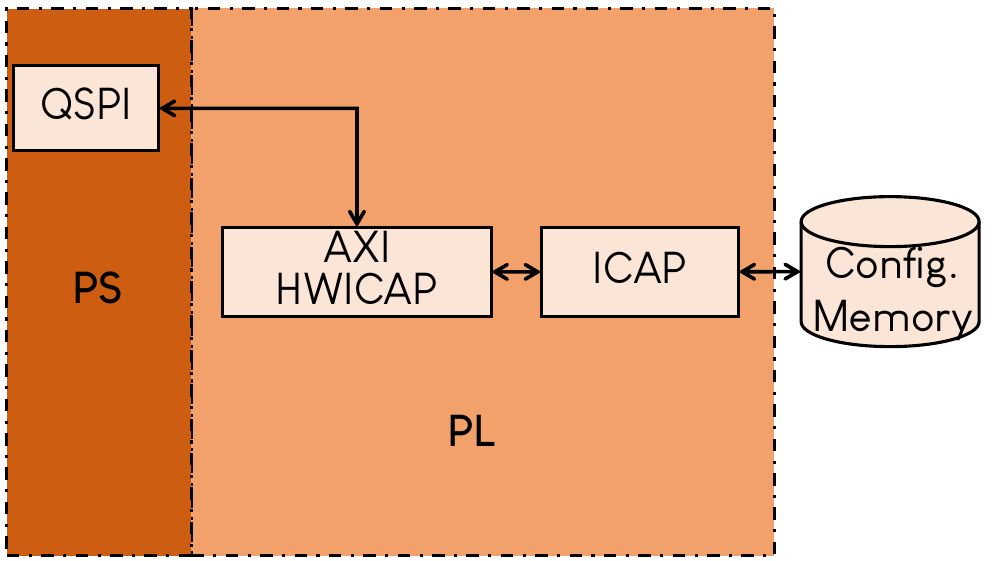}}
\subfloat[TMR\label{z3}]{\includegraphics[width=0.33\textwidth]{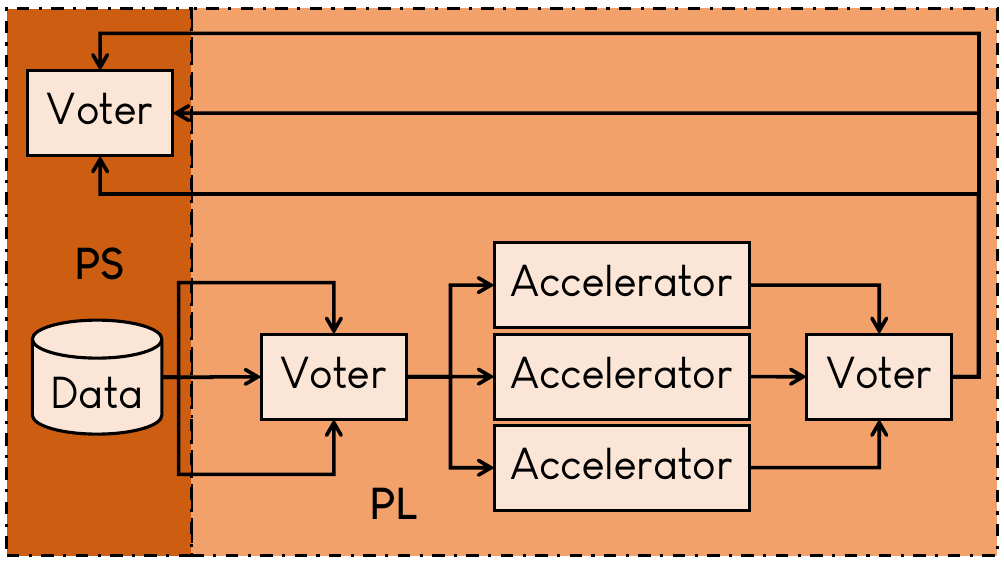}}
\caption{FT techniques on Zynq:
(a) Configuration Memory Scrubbing,
(b) Dynamic Partial Reconfiguration,
and
(c) Triple Modular Redundancy.}
\label{ztech}
\vspace{-12pt}
\end{figure*}

\subsection{CMS: Configuration Memory Scrubbing}
Our design for scrubbing Zynq's configuration memory is based on Xilinx's SEM IP. 
This IP utilizes the internal readback CRC hardware to perform error detection using the built-in ECC and CRC codes,
as shown in Fig. \ref{z1}. 
The Frame ECC interface is used to to access the configuration memory checksum
when a memory frame is read.
The Zynq PS communicates with SEM via the AXI UART interface, 
receiving information about the status of SEM, 
as well as transmitting commands to SEM.
The Data Fetch interface is a mechanism allowing SEM
to fetch data from an external memory,
i.e., the chip's QSPI flash memory in our case, 
and perform memory scrubbing via the ICAP interface. 
In our design,
we configure SEM to ``replace'' mode,
which applies data-reload-based correction of the
configuration memory frames.

\subsection{DPR: Dynamic Partial Reconfiguration}
In case a small part of the design needs to be modified due to errors,
we employ dynamic partial reconfiguration.
As shown in Fig. \ref{z2},
partial bitstreams are delivered to the ICAP via the AXI HWICAP IP core
using the ARM processor and the AXI lite protocol. 
In particular, a software running on ARM 
is responsible for reading bitstream data from the the QSPI flash memory 
and feeding them to ICAP via AXI HWICAP.

\subsection{TMR: Triple Modular Redundancy}
Our third FT technique is the classic TMR,
which, 
however, is tailored to the Zynq architecture
as shown in Fig. \ref{z3}.
Besides only triplicating the main accelerator implemented on the FPGA,
we also applying voting on the input data 
that are transferred to PL from PS via AXI.
Correspondingly,
we apply voting on the PS side to ensure that the correct output data are transferred to ARM.

\subsection{Hybrid Fault-Tolerant Architectures}
At first,
we apply both DPR and CMS (for memory correction)
along with TMR (for application-level protection),
while we also combine all three techniques.
Finally,
to increase the robustness of the system,
we develop an architecture where all the proposed FT techniques are applied in cooperation with a watchdog policy (WD) running on an external microcontroller (prototyped with Arduino).
In particular,
if any of the TMR components are permanently faulty,
we apply partial reconfiguration.
At the same time, the SEM controller performs CMS
and provides a message to the watchdog timer via UART.
In case the timer expires,
we reset the entire SoC FPGA by using the configuration data stored in the QSPI flash memory.
The combination of all the FT techniques imposes some limitations.
Both CMS and DPR use the ICAP peripheral for reconfiguring the FPGA.
Therefore, ICAP is available only to SEM or HWICAP
to apply CMS or DPR, respectively. 
Moreover,
in the WD-based architecture,
we configure SEM to 
``enhanced repair'' mode (ECC and CRC algorithm-based correction),
as QSPI is needed for booting the device.

\section{Mitigation Techniques on Myriad2 VPU}

In the Myriad2 VPU (right side of Fig. \ref{archi2}),
the general-purpose LEON core
is the orchestrator of the entire dataflow,
while it also implements the FT techniques.
The input data are received and stored in the DDR global memory,
and then,
they are transferred via DMA 
to the scratchpad 2MB-size CMX memory,
where they are processed by the 
12 SHAVE cores (VLIW \& SIMD processors). 
We note that the developer manually divides the workload to tasks
and assigns them to SHAVEs. 
Below,
we present the FT techniques for Myriad2.

\begin{figure*}[!t]
\centering
\subfloat[IMR\label{m1}]{\includegraphics[width=0.33\textwidth]{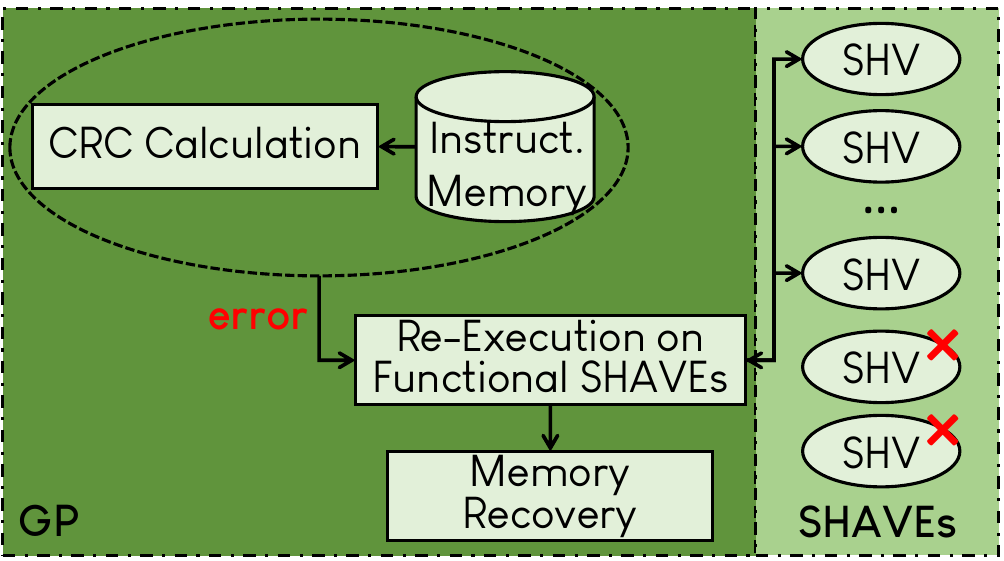}}
\subfloat[DMR\label{m2}]{\includegraphics[width=0.33\textwidth]{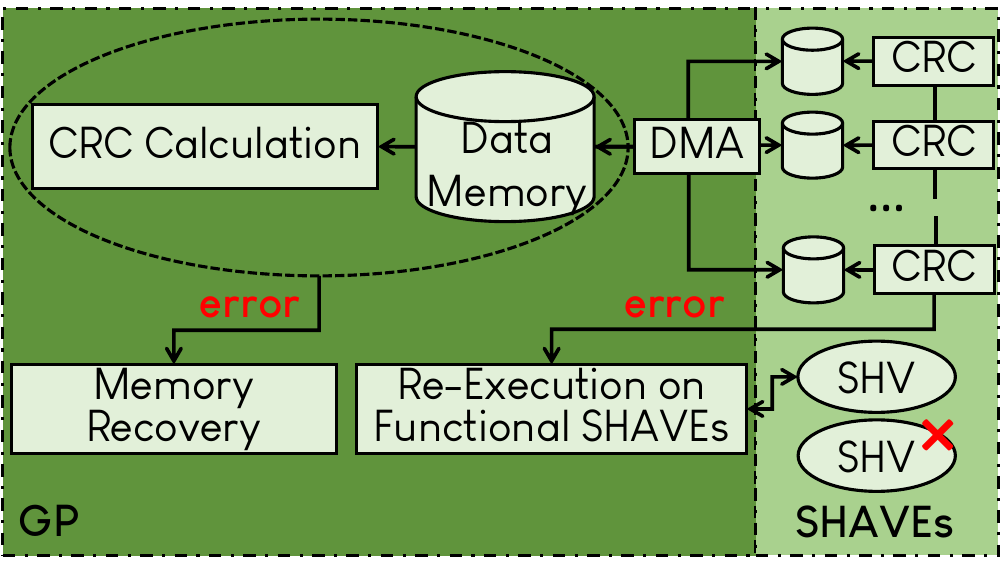}}
\subfloat[NMR\label{m3}]{\includegraphics[width=0.33\textwidth]{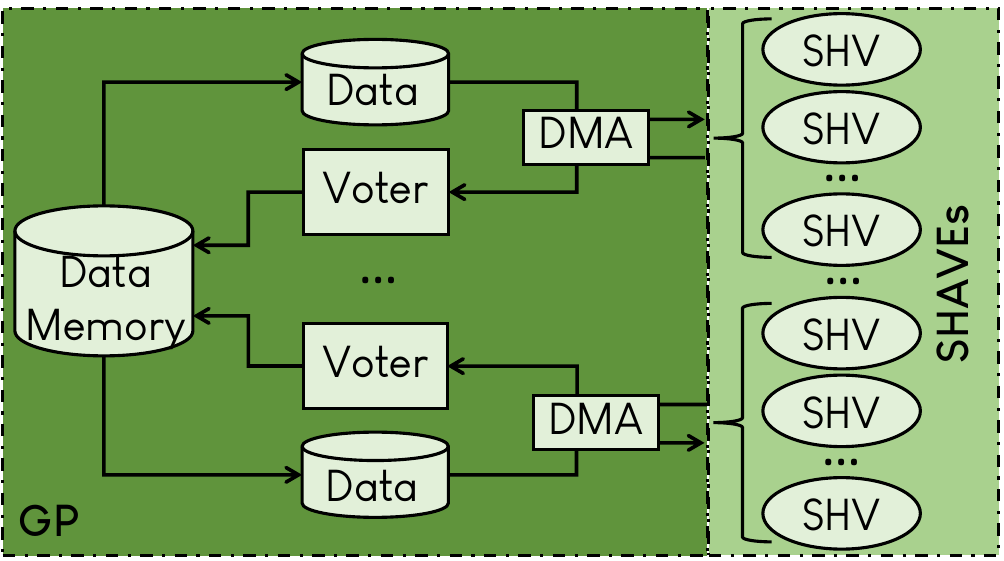}}
\caption{FT techniques on Myriad2:
(a) Instruction Memory Recovery,
(b) Data Memory Recovery,
and
(c) $N$ Modular Redundancy.}
\label{mtech}
\vspace{-8pt}
\end{figure*}

\subsection{IMR: Instruction Memory Recovery}
Our first fault-tolerant technique targets to 
correct errors in the instruction memory of SHAVEs, 
which are the main acceleration cores.
The developer can specify the memory space of the SHAVEs' code 
either in DDR or CMX,
thus, 
LEON knows the instruction memory space of each SHAVE core.  
In this technique,
illustrated in Fig. \ref{m1},
LEON initially calculates the 
CRC values of all the instruction memory space.
Therefore, 
at runtime,
it can check if the memory contents
have changed through the repeated calculation of the CRC.
Considering that 
the SHAVE code is not modified at runtime,
any change in these memory locations indicates fault.
In such case,
LEON detects the SHAVEs corresponding to corrupted memory locations
and sends their input data to the remaining functional SHAVEs
to 
calculate the correct results.
Upon finishing the the re-execution,
LEON recovers the impaired SHAVEs
using a golden copy of the instruction memory
(stored in DDR or external memory). 

\subsection{DMR: Data Memory Recovery}
A similar CRC-based approach,
illustrated in Fig. \ref{m2}, 
is used to confront errors in the data. 
Besides the CRC calculations for the data received via CIF
(presented in Section \ref{comm}),
we detect errors in the data after their on-chip DMA transmission in the CMX scratchpad memory,
where they are directly accessed and processed by SHAVEs.
LEON appends a CRC value to each input data tile,
and the corresponding SHAVE calculates the CRC value before starting the processing. 
In case of errors in the data of some SHAVEs,
LEON restores the input data
(e.g., by using a golden copy such as in the case of the instruction memory),
and then, 
re-schedules the execution on the functional SHAVEs.

\subsection{NMR: $N$ Modular Redundancy}
In this technique,
instead of parallelizing the workload to the 12 SHAVE cores,
we create groups of $N$ SHAVEs and assign them the same input data for processing.
LEON receives their outputs and executes a voting system to retain the correct result.
This process is illustrated in Fig. \ref{m3}.
For $N$=3, we have 4 groups of SHAVEs,
i.e., in practice the workload is parallelized to 4 cores.
For $N$=5, we have 4 groups of SHAVEs,
while we do not use 2 cores at all. 

\section{Fault-Tolerant FPGA--VPU Communication}
\label{comm}

The payload data communication between the FPGA and VPU is performed over the CIF and LCD interfaces in the form of image frames and is designed to be fault-tolerant.
In particular, payload data frames exchanged between both sides are protected by a 16-bit CRC field. 
The CRC values are appended to the end of each frame by means of a frame footer. 
The footer has the form of an entire CIF or LCD frame row with the first pixels containing the 16-bit CRC,
while the rest pixels are zero-padded. 
The CRC algorithm used for both CIF and LCD is CRC-16-CCITT,
where the polynomial is 0x1021 ($x^{\text{16}} + x^{\text{12}} + x^\text{5} + \text{1}$)
with an initial value of 0x0.

\begin{figure}[!t]
\centering
\subfloat[CIF-CRC\label{cifcrc}]{\includegraphics[width=0.49\columnwidth]{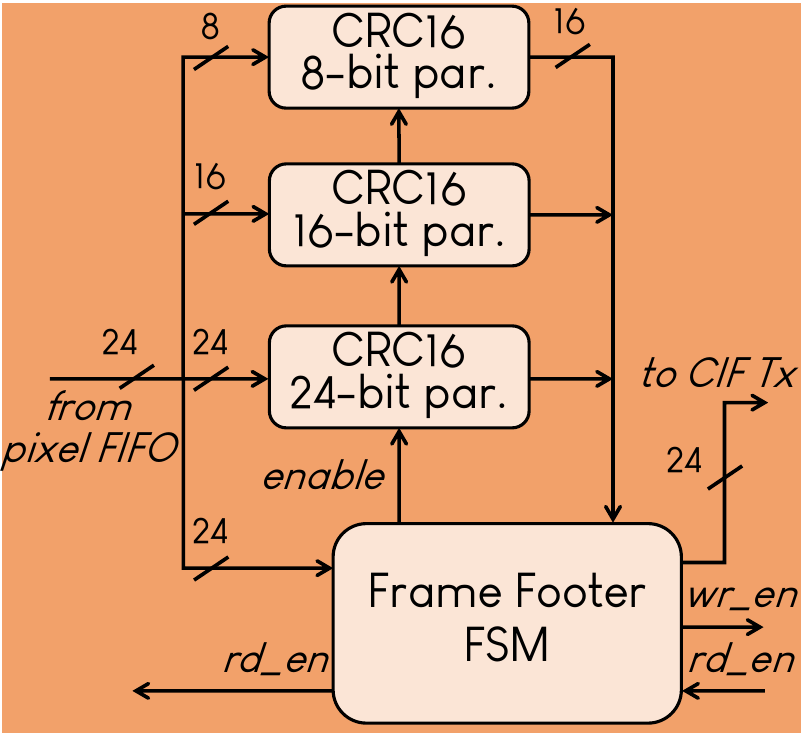}}\hspace{1.5pt}
\subfloat[LCD-CRC\label{lcdcrc}]{\includegraphics[width=0.49\columnwidth]{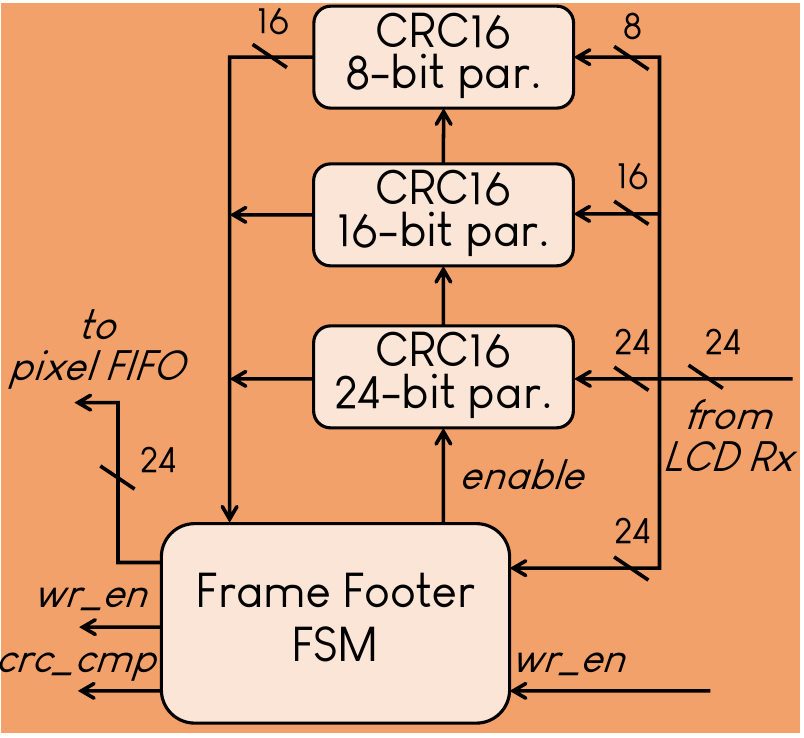}}%
\caption{CRC modules implemented on the FPGA for FT communication.}
\label{ftcomm}
\vspace{-8pt}
\end{figure}

On the FPGA side, the module that is responsible for formulating and appending the CRC on CIF frames transmitted to the VPU is shown in Fig. \ref{cifcrc}. 
While a CIF frame transmission is active, the CIF CRC module reads one pixel value at each clock cycle from a pixel buffer and forwards it to the CIF transmitter. 
The CIF transmitter supports frames with bit-depth of 8, 16 and 24. 
For this, the CIF CRC contains three instances of a CRC16 calculator 
(based on linear-feedback shift register),
which operate in parallel and calculate on-the-fly the CRC values of frames with 8, 16 and 24-bit depth. 
The CRC16 calculators are controlled by the counter-based Frame Footer FSM. When the active portion of the frame has been forwarded, the CRC16 calculation is complete. 
Then, the Frame Footer FSM selects the output of the CRC16 calculators based on the corresponding bit-depth and forwards the additional footer row that contains the CRC16 value in the first pixel(s). 
When the entire CIF frame is received on the VPU side, the VPU calculates the corresponding CRC16 value and compares it with the received one.

The transmission of LCD frames from the VPU to the FPGA follows the same principles as that of CIF frames. 
When an LCD frame of payload data has been formulated in the VPU, 
the VPU calculates the CRC16 value, includes it to the footer of the frame and starts the frame transmission over LCD. 
The module that handles the CRC calculation and comparison on the FPGA side is shown in Fig. \ref{lcdcrc}. 
The LCD receiver feeds the LCD CRC module with one pixel value at each clock cycle. 
Three distinct CRC16 modules operate in parallel on different bit-widths of the incoming pixel in order to support all the three different frame bit-depths. 
When the active portion of the LCD frame has been received, the CRC16 value is calculated and the Frame Footer FSM handles the reception of the additional footer row. 
It extracts the received CRC value from the correct position of the footer and performs the comparison with the calculated one. 
The comparison result is reported to the corresponding status registers of the FPGA. 

\section{Experimental Evaluation}

\subsection{Test Setup}

For the experimental evaluation,
we use the Zybo board (Zynq-7010) 
along with the associated software tools,
i.e., Xilinx's Vivado and SDK, 
and the Myriad2 VPU along with Intel's MDK design suite.
As benchmarks, 
we use an FIR filter for the FPGA
and two image processing software kernels
(averaging binning and floating-point convolution) for the VPU.  
The FIR filter is developed with VHDL on the PL of Zynq,
and applies pipelined filtering for streaming input,
while its coefficients are stored in a ROM mapped onto LUTs. 
The VPU benchmarks
are parallelized to the 12 SHAVE cores,
i.e., 
the input image is divided into stripes that are assigned to SHAVEs for parallel processing. 
The general-purpose LEON processor performs all the necessary high-level tasks,
e.g., SHAVE initialization, stripe partition, and DMA transfers.\\[-9pt]

\noindent\underline{Zynq FPGA}:
To emulate errors in the FPGA, 
 we inject faults in Zynq's configuration memory
using the SEM-based ACME tool \cite{inj_fpga}.
This tool translates the essential memory bits,
which are provided in the so-called EBD ASCII file,
into injection addresses for SEM.
ACME also provides
subsets of essential bits that 
correspond to the components of the design,
allowing to perform injection campaigns for specific design parts.
For injecting faults,
the Zynq PS sends commands to SEM via AXI-UART,
specifying the bit locations in the memory frame addresses. 
The successful fault injection 
is indicated by status signals,
while the SEM controller transitions from the injection state to idle
to wait for the next injection command. \\[-9pt]

\noindent\underline{Myriad2 VPU}:
Intel does not provide any fault-injection framework 
such as SEM,
thus, we develop our own methods for
inserting faults in the data and instructions memories.
Regarding data,
we consider three approaches that are based on random-number routines:
(i) LEON corrupts the input data 
(stored in DDR)
before transferring them to CMX to be processed by SHAVEs,
(ii) SHAVEs corrupt their own input data upon receiving them, 
(iii) LEON corrupts the shared variables between it and SHAVEs.
Similarly,
we assign LEON to corrupt the instructions (code) of SHAVEs (stored in DDR).

\subsection{FPGA Results}

We employ a test involving 10K fault injections inserted in the configuration memory every 4ms.
Our injection campaign targets the utilized FPGA area (FT components + application). 

\begin{figure}[!t]
\vspace{-5pt}
    \centering
    \includegraphics[width=0.91\columnwidth]{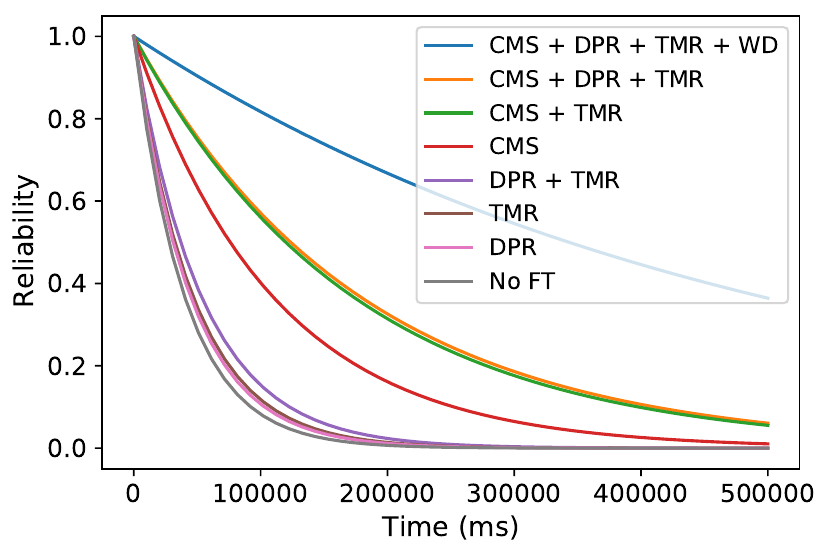}
    \vspace{-7pt}
    \caption{FPGA reliability (failure probability) w.r.t. to the FT techniques (legend sorted based on declining reliability).}
    \label{reli}
    \vspace{-7pt}
\end{figure}

The results of Table \ref{tb_rel}
show the portion of time that the FPGA
is non-functional (down)
and has erroneous/correct functionality. 
As shown,
the standalone application of each FT technique
improves the downtime 
from 92\% to 73--80\%, with CMS exhibiting the correct functionality for more time (22\% vs. 1--8\%). 
The advantage of CMS is that it detects the errors when they generated (it checks the memory and does not wait for a new output, i.e., like TMR).
Furthermore, TMR provides near-zero correct functionality,
as our injection campaign targets the TMR modules
and after the $\sim$2K injections the system is down.
Nevertheless,
when combining TMR with either CMS or DPR,
the functionality of the FPGA is improved. 
On the one hand,
this is due to the errors 
being distributed across a larger design,
but also because TMR notifies in time DPR to act
and increases the time for CMS to correct the errors.
The combination of all three FT techniques
increases the reliability of the system even more,
as the FPGA operates correctly for half of the test duration.
Finally,
the benefit of additionally using WD
is that, besides quickly detecting errors by both TMR and CMS,
it captures the case that the CMS design has a permanent fault.
As a result, 
it significantly increases the robustness of the system,
providing correct execution for 72\% of time.
Based on the same test,
Fig. \ref{reli} demonstrates the reliability
of the FT architectures over time.
We assume exponentially-distributed random faults at rate $\lambda$ and that the
system reliability is defined as
$R(t) = e^{-\lambda \cdot t}$ \cite{Afsharnia17}.
The pattern remains the same,
as the most reliable architectures are the combined one.
However, 
DPR provides less reliability,
as it lacks an error detection mechanism
contrary to the other techniques.

\begin{table}[!t]
\renewcommand{\arraystretch}{1.2}
\setlength{\tabcolsep}{2.5pt}
\caption{FPGA Test: 10K Fault Injections in utilized area (one per 4ms)}
\vspace{-5pt}
\label{tb_rel}
\centering
\begin{threeparttable}
\begin{tabular}{l|ccc}
\hline 
\multicolumn{1}{c|}{\multirow{2}{*}{\textbf{FT Architecture}}} & \textbf{Downtime / No} & \textbf{Erroneous} & \textbf{Correct} \\[-2pt]
 & \textbf{Functionality} & \textbf{Functionality} & \textbf{Functionality} \\
 \hline \hline \\[-8.5pt]
No FT (app only)                      
&   92\% & 8\%& 0\% \\ \hline\\[-9pt]
TMR	                       
&   80\% & 19\% & 1\%\\
DPR                                            
&   79\% & 13\% & 8\%\\
CMS	                                           
&   73\% & 5\%& 22\%\\
DPR \hspace{-1.5pt}+\hspace{-1.5pt} TMR	               
&   36\% & 41\% & 23\%\\
CMS \hspace{-1.5pt}+\hspace{-1.5pt} TMR 	               
&   31\% & 24\% & 45\%\\
CMS \hspace{-1.5pt}+\hspace{-1.5pt} DPR \hspace{-1.5pt}+\hspace{-1.5pt} TMR 	       
&   27\% & 22\% & 51\%\\
CMS \hspace{-1.5pt}+\hspace{-1.5pt} DPR \hspace{-1.5pt}+\hspace{-1.5pt} TMR \hspace{-1.5pt}+\hspace{-1.5pt} WD   
&   20\% & 8\% & 72\%\\
\hline
\end{tabular}
\begin{tablenotes}
   \item[1]{\fontsize{6.5}{7.7}\selectfont \% refers to portion of time of the total test duration.}
   \end{tablenotes}
 \end{threeparttable}
 \vspace{-3pt}
\end{table}
\begin{table}[!t]
\renewcommand{\arraystretch}{1.2}
\setlength{\tabcolsep}{6.5pt}
\caption{Resource Utilization of FT Techniques on Zynq-7010}
\vspace{-5pt}
\label{tb_ftfpga}
\centering
\begin{tabular}{l|ccc}
\hline 
\multicolumn{1}{c|}{\textbf{FT Architecture}} & \textbf{LUTs} & \textbf{FFs}  & \textbf{RAMBs} \\
 \hline \hline \\[-8.5pt]
No FT (app only) 		                       &   609 (4\%) &  1461 (4\%) &  1	 (2\%)  \\\hline\\[-9pt]
TMR	                       &   1316	(8\%) & 2327 (7\%) & 	4 (7\%) \\
DPR                        &   1315 (8\%) & 2501 (7\%) &     0  (0\%)  \\
CMS	                       &   2021 (12\%) & 2761 (8\%) &  	2 (3\%)  \\
DPR \hspace{-1.5pt}+\hspace{-1.5pt} TMR	               &   2022	(12\%) & 3367 (10\%) &	4 (7\%) \\
CMS \hspace{-1.5pt}+\hspace{-1.5pt} TMR 	               &   2727 (16\%) & 3801 (11\%) &		6 (10\%) \\
CMS \hspace{-1.5pt}+\hspace{-1.5pt} DPR \hspace{-1.5pt}+\hspace{-1.5pt} TMR 	       &   3433 (20\%) & 4841 (14\%) & 		6 (10\%) \\
CMS \hspace{-1.5pt}+\hspace{-1.5pt} DPR \hspace{-1.5pt}+\hspace{-1.5pt} TMR \hspace{-1.5pt}+\hspace{-1.5pt} WD   &   3756	(21\%) & 5072 (14\%) &		6 (10\%) \\
\hline
\end{tabular}
\vspace{-7pt}
\end{table}

Table \ref{tb_ftfpga} shows that
all the FT architectures exhibit small resource utilization ($<$4K LUTs, $<$5K FFs, $<$6 RAMBs). 
Therefore, there is room for the implementation of other components required in the co-processing architecture, e.g., more DSP functions, data transcoders and data compressors.
In more detail,
the SEM IP of the CMS technique
imposes the largest resource overhead. 
The resource increment in the WD-based FT architecture is due to: 
(i) the UART communication between
the watchdog (microcontroller) and the FPGA,
and (ii)
the ``enhanced repair'' mode of SEM 
(instead of ``replace'').
Finally,
according to our measurements,
SEM corrects a memory frame of 404 bytes
in 18ms (``enhanced repair''),
while the the throughput of DPR via HWICAP is 67 MB/s.

\subsection{VPU Results}
For the Myriad2 VPU,
we examine the error rate,
i.e., the number of erroneous pixels per total pixels, 
as well as the timing overheads
for providing FT processing.
Our tests involve the random injection of errors
in the instruction and data memories of 3, 6, 9, or 12 SHAVE cores.

Table \ref{tb_m2} reports the range 
of the error rate for all our tests without and with the FT techniques.
The convolution kernel provides less error resilience than averaging binning,
however, as expected,
their error rate reaches almost 100\% 
when corrupting the memories of all 12 SHAVEs.
The application of either IMR or DMR,
i.e., the CRC-based FT techniques 
that reschedule the workload of impaired SHAVEs,
eliminates the errors.
Regarding the standalone application of NMR
i.e., without rescheduling, 
most of the errors remain,
especially when corrupting $\geq$6 SHAVEs
(because 2 or more cores of the same voting group are impaired).

\begin{figure}[!t]
    \centering
    \includegraphics[width=0.85\columnwidth]{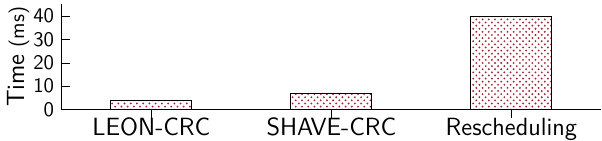}
    \vspace*{-9pt}
    \caption{Overhead of main tasks for FT processing on Myriad2 VPU.}
    \label{m2t}
    \vspace{-7pt}
\end{figure}

In terms of timing overhead,
NMR is equivalent to applying
4-core ($N$=3) and 2-core ($N$=5)
rather than 12-core parallelization.
Namely,
there is an almost linear decrease in the
application latency
(when also considering the voting overhead).
For IMR and DMR,
Fig. \ref{m2t} shows the execution times of their main tasks.
The CRC calculations are small ($<$10ms), 
while the rescheduling (data re-arrangement and assignment to non-impaired SHAVEs)
requires $\sim$40ms on average.
Nevertheless, both techniques achieve correct application execution.

\begin{table}[!t]
\renewcommand{\arraystretch}{1.2}
\setlength{\tabcolsep}{4pt}
\caption{Error Rate of FT Processing on Myriad2 VPU}
\label{tb_m2}
\vspace{-5pt}
\centering
\begin{threeparttable}
\begin{tabular}{cl|cccc}
\hline 
 \multicolumn{2}{c|}{\multirow{2}{*}{\textbf{Benchmark}}} & \multicolumn{4}{c}{\textbf{Impaired SHAVE Cores}}\\
& & \textbf{3}  & \textbf{6} & \textbf{9} & \textbf{12} \\
 \hline \hline 
\multirow{2}{*}{No FT (app only)} & Conv2D &  13--25\%&  24--50\% & 31--74\% &  44--99\% \\
& Binn2D &  9--25\% & 12--48\% &  16--74\% & 24--97\% \\
\hline
\multirow{2}{*}{IMR \hspace{0pt}$\mid$\hspace{0pt} DMR \hspace{0pt}$\mid$\hspace{0pt} NMR} & Conv2D &  0--13\%&  0--48\% & 0--74\% &  0--99\% \\
& Binn2D &  0--11\% & 0--48\% &  0--69\% & 0--97\% \\
\hline
\end{tabular}
\begin{tablenotes}
   \item[1]{\fontsize{6.5}{7.7}\selectfont Impaired SHAVE refers to faults in its data or instruction memory.}
   \item[2]{\fontsize{6.5}{7.7}\selectfont The errors in FT processing remain in the NMR technique.}
   \end{tablenotes}
 \end{threeparttable}
 \vspace{-5pt}
\end{table}

\subsection{Communication Results}

Regarding the FPGA–VPU communication, 
the VHDL design for the CIF transmitter and LCD receiver, including the CRC modules, can achieve an operating frequency of 100MHz on the PL
of Zynq-7010.
The resource utilization for both the receiver and transmitter as well as the additional CRC modules is shown in Table \ref{tb_ftcomm}. 
We observe that the fault-tolerance CRC functionality of the communication adds 
small
resource overhead,
which is similar to that of the FT techniques. 

Given that the VHDL can support the frequency of 100MHz, the actual operating frequency and achievable throughput for reliable communication depends on the I/O capabilities of different hardware setups. 
For example, in commercial hardware, a Xiphos Q7S board with a custom mezzanine daughter card connected to an Eyes Of Things (EoT) Myriad2 board over jumper wires can support reliable transmission over CIF for frames of 1024$\times$1024, 16-bit pixels at 50MHz. The same connectivity principles can be applied to the space-oriented COTS Myriad2 CogniSat platform. 
The FPGA-VPU communication can scale up to both CIF \& LCD operating reliably up to 150MHz for 2048$\times$2048, 24-bit frames in custom hardware setups for space applications, 
where the FPGA and Myriad2 are connected over a custom FMC Mezzanine connector as in \cite{OBDP_Navarro} and \cite{ICECS_HPCB}.

\begin{table}[!t]
\renewcommand{\arraystretch}{1.2}
\setlength{\tabcolsep}{4.5pt}
\caption{Resources for FT FPGA--VPU Communication on Zynq-7010}
\vspace{-5pt}
\label{tb_ftcomm}
\centering
\begin{tabular}{l|cccc}
\hline 
\multicolumn{1}{c|}{\textbf{Component}} & \textbf{LUTs} & \textbf{FFs}  & \textbf{DSPs} &\textbf{RAMBs} \\
 \hline \hline 
CIF Tx \hspace{-1.5pt}+\hspace{-1.5pt} LCD Rx                      &   2977 (17\%) & 1232 (4\%) &  2 (3\%)    &	10 (17\%) \\
CIF-CRC \hspace{-1.5pt}+\hspace{-1.5pt} LCD-CRC                        &   347 (2\%) & 305 (1\%) &  2 (3\%)    &    0 (0\%)   \\
\hline
\end{tabular}
\vspace{-9pt}
\end{table}

\section{Conclusion}
We developed, proposed, and combined various FT techniques
for the Zynq FPGA and the Myriad2 VPU,
i.e., two COTS chips already being considered for space missions.
We prototyped 
an FPGA \& VPU co-processing architecture based on available COTS boards,
we evaluated techniques
for error correction in memories,
accelerator redundancy, watchdog timers, 
and data transmission based on CRC. 
We showed that
FPGA reliability gets significantly increased
via our robust hybrid FT architecture,
which utilizes $\sim$4K LUTs
to provide correct functionality
for 72\% of the time during very demanding tests.
Regarding Myriad2, 
we achieve error-free execution
with an overhead of up to $\sim$50ms.
Future work will include engineering models
with space-heritage components.

\bibliographystyle{IEEEtran}
\bibliography{REFERENCES.bib}

\end{document}